\documentclass[reprint,superscriptaddress,amsmath,amssymb,aps,prb,showpacs]{revtex4-2}
\usepackage{lineno}
\usepackage[english]{babel}
\usepackage{multirow}
\usepackage[thinspace]{SIunits}
\usepackage{xspace}
\usepackage[normalem]{ulem}
\usepackage{soul}
\usepackage{graphicx}
\graphicspath{{figures//}} 
\usepackage{dcolumn}
\usepackage{bm}
\usepackage{color}
\usepackage{wrapfig}
\usepackage{hyperref}
\usepackage{array}
\usepackage{booktabs}
\usepackage{multirow}

\hypersetup{
    unicode=false,          
    pdftoolbar=true,        
    pdfmenubar=true,        
    pdffitwindow=false,     
    pdfstartview={FitH},    
    pdftitle={My title},    
    pdfauthor={Author},     
    pdfsubject={Subject},   
    pdfcreator={Creator},   
    pdfproducer={Producer}, 
    pdfkeywords={keyword1} {key2} {key3}, 
    pdfnewwindow=true,      
    colorlinks=true,       
    linkcolor=blue,          
    citecolor=blue,        
    filecolor=blue,      
    urlcolor=blue,       
    plainpages=false        
}

\newcommand{\Tc}{$T_{\mathrm{{C}}}$\xspace}

\newcommand{\CSS}{\mbox{Co$_{3}$Sn$_{2}$S$_{2}$}\xspace}

\newcommand{\YBCO}{\mbox{YBa$_{2}$Cu$_{3}$O$_{7}$}\xspace}
\newcommand{\Alg}{$A_{1\rm g}$\xspace}

\newcommand{\AZg}{$A_{2\rm g}$\xspace}
\newcommand{\Blg}{$\rm{B_{1g}}$\xspace}

\newcommand{\Eg}{$E_{\rm g}$\xspace}
\newcommand{\Egl}{$E_{\rm g}^{(1)}$\xspace}
\newcommand{\Egz}{$E_{\rm g}^{(2)}$\xspace}
\newcommand{\grd}{$^{\circ}$\xspace}

\newcommand{\wn}{$\rm{cm}^{-1}$\xspace}

\begin{document}

\title{ \Large  Chiral electronic excitations and strong electron-phonon coupling to Weyl fermions in the Kagome  semimetal Co$_3$Sn$_2$S$_2$}

\author{Ge He}
\altaffiliation{These authors contributed equally to the work.}
\affiliation{School of Mechanical Engineering, Beijing Institute of Technology, Beijing 100081, China}
\affiliation{Walther Meissner Institut, Bayerische Akademie der Wissenschaften, Garching 85748, Germany}
\affiliation{Department of Physics, University College Cork, College Road, Cork T12 K8AF, Ireland}

\author{Malhar Kute}
\altaffiliation{These authors contributed equally to the work.}
\affiliation{Department of Materials Science and Engineering, Stanford University, Stanford, California 94305, USA}
\affiliation{Stanford Institute for Materials and Energy Sciences, SLAC National Accelerator Laboratory and Stanford University,
2575 Sand Hill Road, Menlo Park, California 94025, USA}

\author{Zhongchen Xu}
\affiliation{Beijing National Laboratory for Condensed Matter Physics, Institute of Physics, Chinese Academy of Sciences, Beijing 100190, China}
\affiliation{School of Physical Sciences, University of Chinese Academy of Sciences, Beijing 100049, China}

\author{Leander Peis}
\altaffiliation{Present address: Capgemini, Frankfurter Ring 81,\newline 80807 M\"unchen, Germany}
\affiliation{Walther Meissner Institut, Bayerische Akademie der Wissenschaften,
Garching 85748, Germany}
\affiliation{School of Natural Sciences, Technische Universit\"at M\"unchen, Garching 85748, Germany}
\affiliation{IFW Dresden, Helmholtzstrasse 20, Dresden 01069, Germany}

\author{Ramona Stumberger}
\altaffiliation{Present address: Robert Bosch GmbH, Robert-Bosch-Campus 1,\newline 71272 Renningen, Germany}
\affiliation{Walther Meissner Institut, Bayerische Akademie der Wissenschaften,
Garching 85748, Germany}
\affiliation{School of Natural Sciences, Technische Universit\"at M\"unchen, Garching 85748, Germany}

\author{Andreas Baum}
\altaffiliation{Present address: Mynaric, Bertha-Kipfm\"uller-Str. 2-4,\newline 81249 M\"unchen, Germany }
\affiliation{Walther Meissner Institut, Bayerische Akademie der Wissenschaften, Garching 85748, Germany}
\affiliation{School of Natural Sciences, Technische Universit\"at M\"unchen, Garching 85748, Germany}

\author{Daniel Jost}
\affiliation{Walther Meissner Institut, Bayerische Akademie der Wissenschaften,
Garching 85748, Germany}
\affiliation{Stanford Institute for Materials and Energy Sciences, SLAC National Accelerator Laboratory and Stanford University,
2575 Sand Hill Road, Menlo Park, California 94025, USA}
\affiliation{School of Natural Sciences, Technische Universit\"at M\"unchen, Garching 85748, Germany}

\author{Emily Been}
\affiliation{Stanford Institute for Materials and Energy Sciences, SLAC National Accelerator Laboratory and Stanford University,
2575 Sand Hill Road, Menlo Park, California 94025, USA}
\affiliation{Department of Physics, Stanford University, Stanford, California 94305, USA}
\affiliation{Physical Sciences Division, College of Letters and Science, University of California Los Angeles, Los Angeles, California 90095, USA}

\author{Brian Moritz}
\affiliation{Stanford Institute for Materials and Energy Sciences, SLAC National Accelerator Laboratory and Stanford University,
2575 Sand Hill Road, Menlo Park, California 94025, USA}

\author{Jun Shen}
\affiliation{School of Mechanical Engineering, Beijing Institute of Technology, Beijing 100081, China}

\author{Youguo Shi}
\affiliation{Beijing National Laboratory for Condensed Matter Physics, Institute of Physics, Chinese Academy of Sciences, Beijing 100190, China}
\affiliation{Songshan Lake Materials Laboratory, Dongguan, Guangdong 523808, China}

\author{Thomas P. Devereaux}
\affiliation{Department of Materials Science and Engineering, Stanford University, Stanford, California 94305, USA}
\affiliation{Stanford Institute for Materials and Energy Sciences, SLAC National Accelerator Laboratory and Stanford University,
2575 Sand Hill Road, Menlo Park, California 94025, USA}
\affiliation{Geballe Laboratory for Advanced Materials, Stanford University, Stanford, California 94305, USA}

\author{Rudi Hackl}
\altaffiliation{To whom all correspondence should be addressed: ge.he@bit.edu.cn; ygshi@iphy.ac.cn; tpd@stanford.edu; hackl@tum.de}
\affiliation{Walther Meissner Institut, Bayerische Akademie der Wissenschaften, Garching 85748, Germany}
\affiliation{School of Natural Sciences, Technische Universit\"at M\"unchen, Garching 85748, Germany}
\affiliation{IFW Dresden, Helmholtzstrasse 20, Dresden 01069, Germany}

\date{\today}
\begin{abstract}
We present results of a Raman scattering study of the Kagome ferromagnet \CSS, with a focus on electronic and phononic excitations and their interplay. We provide a theoretical analysis of the electronic band structure, enabling a semi-quantitative explanation of the spectra. A prominent feature in the electronic spectra is a redistribution of spectral weight from low to high energies  in all polarization configurations starting at the Curie temperature \Tc. In the symmetry-resolved spectra, the suppression of the \Alg continuum in the ferromagnetic state arises from the redistribution of electronic states below \Tc, while a strong enhancement of the \AZg continuum is linked to the dynamics of fermions near the Fermi level $E_{\rm F}$ being characterized by spin-momentum locking near Weyl points. The \Alg phonon modulates the position of these Weyl points and couples strongly to the related fermions close to $E_{\rm F}$. These results allow a comprehensive understanding of the bulk band structure evolution as a function of temperature in \CSS, offering key insights for further studies of the driving force behind the long-range magnetic order and novel topological states in this compound.

\end{abstract}


\maketitle

\section{Introduction}

Quasi two-dimensional (2D) materials with transition metals occupying Kagome lattice sites display various ordering instabilities; and thus, they have attracted considerable attention in recent years \cite{Normand:2009, Kang:2020, Yin:2018, Yin:2020, Ortiz:2020}. Nonmagnetic CsV$_3$Sb$_5$,  for instance, exibits charge density wave (CDW) order below $T_{\rm CDW}=95$~K, and becomes superconducting below $T_{\rm c}=2.5$~K \cite{Ortiz:2020}. \CSS orders ferromagnetically  with the Co spins pointing out of the Kagome plane at low temperature but lying almost entirely in the plane and forming a nearly frustrated antiferromagnetic (AF) pattern close to $T_{\rm C}=175$~K \cite{Guguchia:2020}. 

The generic band structure of these compounds, combining two intersecting bands resulting from the graphene-like hexagons and a flat band from the triangles \cite{Kiesel:2012}, is believed to contribute to these properties. Without any additional interaction the intersection points are massless Dirac points such as in graphene. When spin-orbit coupling is switched on the dispersing bands start interacting, and gaps open up. If inversion symmetry (IS), time reversal symmetry (TRS)- such as in the case of ferromagnetic order, or both are broken, Weyl fermions are expected to appear \cite{Armitage:2018}  with spin-momentum locking. 

\CSS is a prototypical ferromagnetic Weyl semimetal. The Fermi arcs expected for Weyl electrons were observed by scanning tunneling microscopy (STM) and angle-resolved photoemission spectroscopy (ARPES) \cite{Liu:2019,Morali:2019,Liu:2019}. The large intrinsic anomalous Hall effect \cite{Wang:2018,Liu:2018}, the zero-field Nernst effect \cite{Guin:2019}, the redistribution of spectral weight below \Tc found by infrared spectroscopy \cite{Yang:2020, Xu:2020}, Faraday rotation, and the magneto-optical Kerr effect \cite{Okamura:2020} are consistent with this view. Although parts of these observations may be traced back to band structure effects, such as transitions between the Co $3d$ orbitals \cite{Yang:2020}, Okamura and coworkers \cite{Okamura:2020} argue that the main contribution originates from the topological nature of the electrons around the Weyl points, which are dominated by the effects of Berry curvature and spin-momentum locking. While this interpretation qualitatively agrees with theoretical predictions, it is highly desirable to probe more directly the dynamics of Weyl fermions and their evolution across the ferromagnetic transition.

Raman spectroscopy affords us the opportunity to address this question in different ways. We can look at the chiral excitations directly in the \AZg channel \cite{Hsu:2025} or we can look for strongly coupled phonons that move the position of the Weyl points. Here, a detailed study of specific phonons reveals new insights such as the temperature dependence of Fano line shapes \cite{Breit:1936, Fano:1961} as studied successfully in many materials \cite{Klein:1982, Devereaux:1995}. Studies in materials with Dirac points or Weyl nodes such as graphene \cite{Kuzmenko:2009, Castro:2007, Li:2012}, TaAs \cite{Xu:2017}, or LaAlSi \cite{Zhang:2020} have revealed substantial line asymmetries for phonons indicative of strong electron-phonon interactions.

\begin{figure*}[ht!]
  \centering
  \includegraphics[width=17cm]{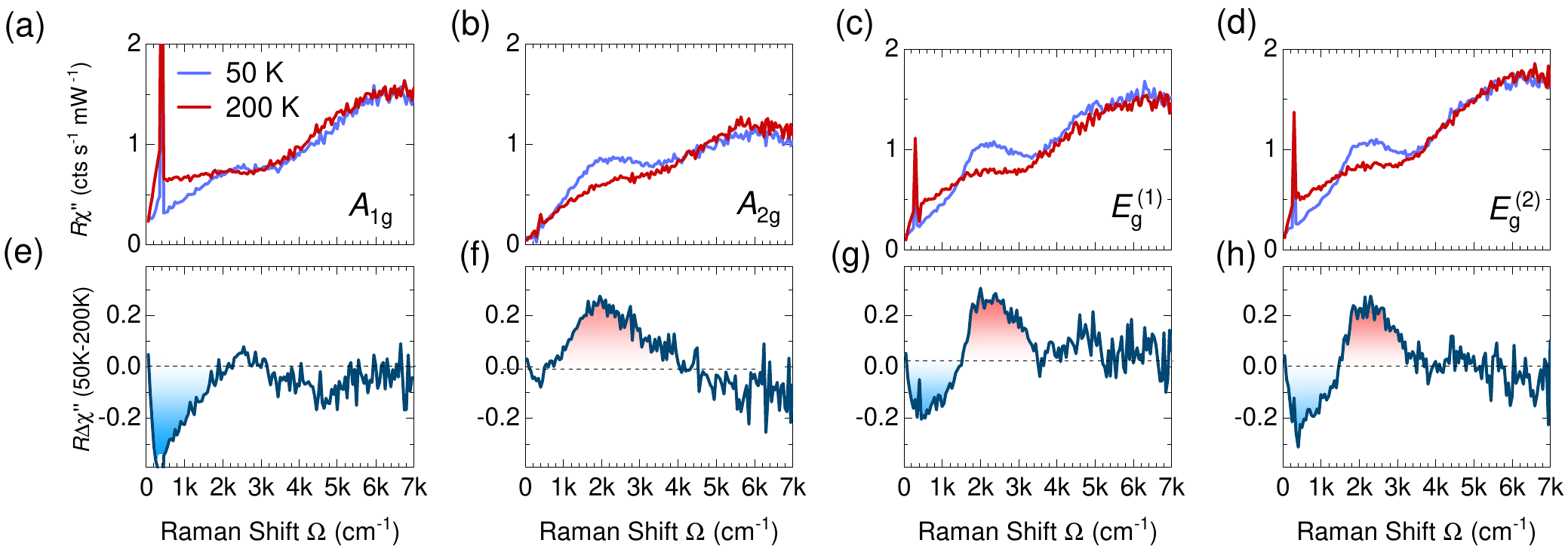}
   \caption{\textbf{Electronic Raman response of \CSS. } (a-d) Raman spectra at 50~K and 200~K in \Alg, \AZg, \Egl and \Egz symmetries, respectively. The measurements were performed with the blue line of the Ar laser ($\lambda_L =476$~nm). (e-h) Corresponding difference spectra. The 200-K spectra has been subtracted from the 50-K spectra.}
	\label{fig:pure_symmetry}
\end{figure*}

In this paper, we present a  Raman study on high-quality single crystals of \CSS at temperatures ranging from 14\,K to 310\,K and determine
all symmetries accessible in the Kagome plane. Two Raman active phonons are identified having differing coupling strengths to the electrons.  The electronic continuum exhibits a spectral weight redistribution below \Tc  with a loss of spectral weight below 1200\,\wn along with an increase for higher energies. The temperature dependences of the \Alg phonon and  continuum argue in favor of electron-phonon coupling (EPC)  originating from the modulation of  the crystal field along the $c$-axis by the motion of the sulfur atoms.  This motion  induces the Weyl points to move in reciprocal space.  The non-monotonic coupling strength for the \Alg phonon and the redistribution of the spectral weight  indicate that bands split and shift across \Tc. Additionally, our simulations of the resonant \AZg response suggest the presence of spin-flip excitations between the Weyl bands. 

\section{Methods}

\subsection{Samples}
\CSS single crystals were grown by a self-flux method as described in Ref. \cite{Liu:2018}. The samples were characterized by X-ray diffraction (XRD) and magnetization measurements, showing a high crystallinity and a ferromagnetic phase transition at \Tc$ = 175$\,K (see more details in Supplemental Information (SI) Section A).

\subsection{Raman experiments}

The experiments were performed with a scanning spectrometer with the resolution set at 4.5~\wn at 575~nm. Since one of the phonon lines has a width comparable to the resolution, a Voigt function (convolution of a Gaussian and a Lorentzian) is used for describing the phonons. For excitation a solid-state laser emitting at 575~nm (Coherent GENESIS MX-SLM577-500) was  incident on the sample at an angle  of 66\grd. Several experiments were performed with the green (514~nm), blue (476~nm) and violet (458~nm) lines of an Ar laser. The incoming light was polarized outside the cryostat in a way that the polarization inside the sample had the desired components. The sample was mounted with good thermal contact to the cold finger of a $^4$He-flow cryostat with the surface normal parallel to the optical axis of the spectrometer. The local heating by the laser was determined to be approximately 1~K/mW absorbed power $P_{\rm a}$. For most of the experiments $P_{\rm a}$ was set at 2~mW. The vacuum was better than $10^{-6}$~mbar.

The Raman susceptibilities shown in the figures, $R\chi^{\prime\prime}(\Omega, T)$, are equal to the dynamical structure factor $S(\Omega, T)$ divided by the Bose-Einstein distribution function $\{1+n(\Omega, T)\}$, where $R$ is an experimental constant, and $S(\Omega,T)$ is proportional to the rate of scattered photons \cite{Devereaux:2007}. Since \CSS has a hexagonal lattice [point group $D_{3d}$, space group $R{\bar{3}}m$ or 166, see also Fig.~\ref{fig:A2g}~(a)], circular polarizations are convenient and were used in most of the experiments. $\hat{R}\hat{R}$ ($\hat{R}\hat{L}$) means that incoming and outgoing photons were both right-circularly (right- and left-circularly) polarized etc., where $\hat{R},\hat{L} = \hat{x}\pm i\hat{y}$ with the perpendicular linear polarizations $\hat{x}$ and $\hat{y}$. The position of the atoms and related space group of \CSS entail the existence of two Raman active modes having \Alg and \Eg symmetry. They appear in the spectra measured with $\hat{R}\hat{R}$ and $\hat{R}\hat{L}$ polarization configurations, respectively, in the Kagome Co$_3$Sn plane (see SI Fig.~S2). In addition to \Alg , \AZg excitations are projected in $\hat{R}\hat{R}$ configuration and are relevant for the electronic or spin response. The details of the polarization configurations and Raman selection rules can be found in the SI Sections B and C.

\subsection{DFT calculations}

Electronic and phonon properties were calculated using density functional theory (DFT) as implemented in the Vienna Ab Initio Simulation Package (VASP) \cite{Kresse:1993, Kresse:1996}, using projector augmented wave (PAW) pseudo-potentials and the Perdew-Burke-Ernzerhof (PBE) functional \cite{Perdew:1996}. The calculations included spin-orbit coupling (SOC) in both the para\-magnetic (PM) and the ferromagnetic (FM) state with  an energy cut-off of 500~eV. Self-consistent field calculations were performed on a 7$\times$7$\times$7 \textbf{k}-point grid, which was increased to 39$\times$39$\times$39 for density of states calculations and reduced to 3$\times$3$\times$3 for phonon supercell calculations. The structures were relaxed independently for the ferromagnetic and paramagnetic phases such that the forces on atoms were less than $10^{-7}$ eV/{\AA}. The phonon energies and eigenvectors were calculated using the Phonopy package \cite{phonopy-phono3py-JPCM,phonopy-phono3py-JPSJ}. Raman-active phonons were identified at the $\Gamma$ point corresponding to the \Alg and \Eg phonons found experimentally.

The electron-phonon coupling (EPC) for the Raman-active phonons was estimated using the frozen phonon method \cite{cohen1990theoretical}. Atoms were displaced along the phonon eigenvectors, and the electron-phonon matrix element was calculated as follows,
\begin{equation}\label{eq:g_epc}
    g_{i, \lambda}(k) = \sqrt{\frac{\hbar}{2m_{\rm S} \omega}} \frac{\partial E_i}{\partial x}
\end{equation}
where $i$ is the band index, $\lambda$ is the phonon index, $m_{\rm S}$ is the mass of sulfur (the only moving atom in this case), $\omega$ is the phonon angular frequency, and $\partial E / \partial x$ is the change  in band energy from the unperturbed case  with respect to the displacement of S. From this, we are able to calculate the overall EPC constant as
\begin{equation}\label{eq:lambda_EPC}
    \lambda_{\rm EPC} = \frac{N_{\rm F} g^2}{\hbar \omega},
\end{equation}
where $N_{\rm F}$ is the density of states at the Fermi level for both spins, and $g$ is the EPC matrix element averaged over the Fermi  momentum  \cite{Devereaux:1995}.

The simulation of the electronic response includes resonant and nonresonant contributions \cite{Shastry:1990}. Transitions with chiral excitations have \AZg ($xy-yx$) symmetry and are not observable off-resonance \cite{Ko:2010}. They must be treated in the presence of intermediate states of proper symmetry at high energies.  Excitations in the Raman active channels (\Alg, \Eg) may either originate from the low-energy interband transitions or by intraband transitions of interacting electrons, for instance in the presence of elastic (defects) or inelastic (phonons, magnons etc.) processes.

We created a tight-binding model from our band structure with Maximally-Localized Wannier Functions (MLWFs) using Wannier90 \cite{wannier90}, with $d_{xy}$, $d_{x^2-y^2}$, and $d_{z^2}$ orbitals on the Co atoms as the initial projections. This model was used to calculate the resonance contribution to the Raman spectra via intermediate states using the Kramers-Heisenberg formula. The details of this calculation can be found in SI Sections J and K.

\section{Results}
\subsection{Electronic excitations}
\label{subsec:electrons}

The electronic Raman response in all symmetries is presented in Fig.~\ref{fig:pure_symmetry}. The procedures for extracting these spectra reproducibly for different laser lines can be found in SI Sections C and E. In all cases, a redistribution of the spectral weight is observed below \Tc, although the behavior varies across different channels. A suppression of the spectral weight is evident in the \Alg symmetry below 1500~\wn [Fig.~\ref{fig:pure_symmetry} (a) and (e)]. In contrast, the \AZg response shows an enhancement of spectral weight between 1000 and 4000~\wn [Fig.~\ref{fig:pure_symmetry} (b) and (f)]. The two degenerate \Eg spectra (\Egl and \Egz) do not show significant differences between each other, with both symmetries exhibiting a small loss of spectral weight at low energies and an increase in spectral weight between 1500~\wn and 4000~\wn [Fig.~\ref{fig:pure_symmetry} (c), (d), (g), and (h)]. As the temperature approaches the Curie point \Tc, the redistribution of spectral weight gradually weakens and finally disappears in all symmetries (see SI Section F for details). In other words, the $E_g$ excitations are qualitatively similar at high and low temperatures, apart from a rearrangement of spectral weight, while $A_{1g}$ and $A_{2g}$ are qualitatively and quantitatively of different nature below $T_C$.
These observations, particularly in the \AZg channel, suggest the presence of a novel type of excitations.\\

\begin{figure*}[ht]
  \centering\includegraphics[width=17cm]{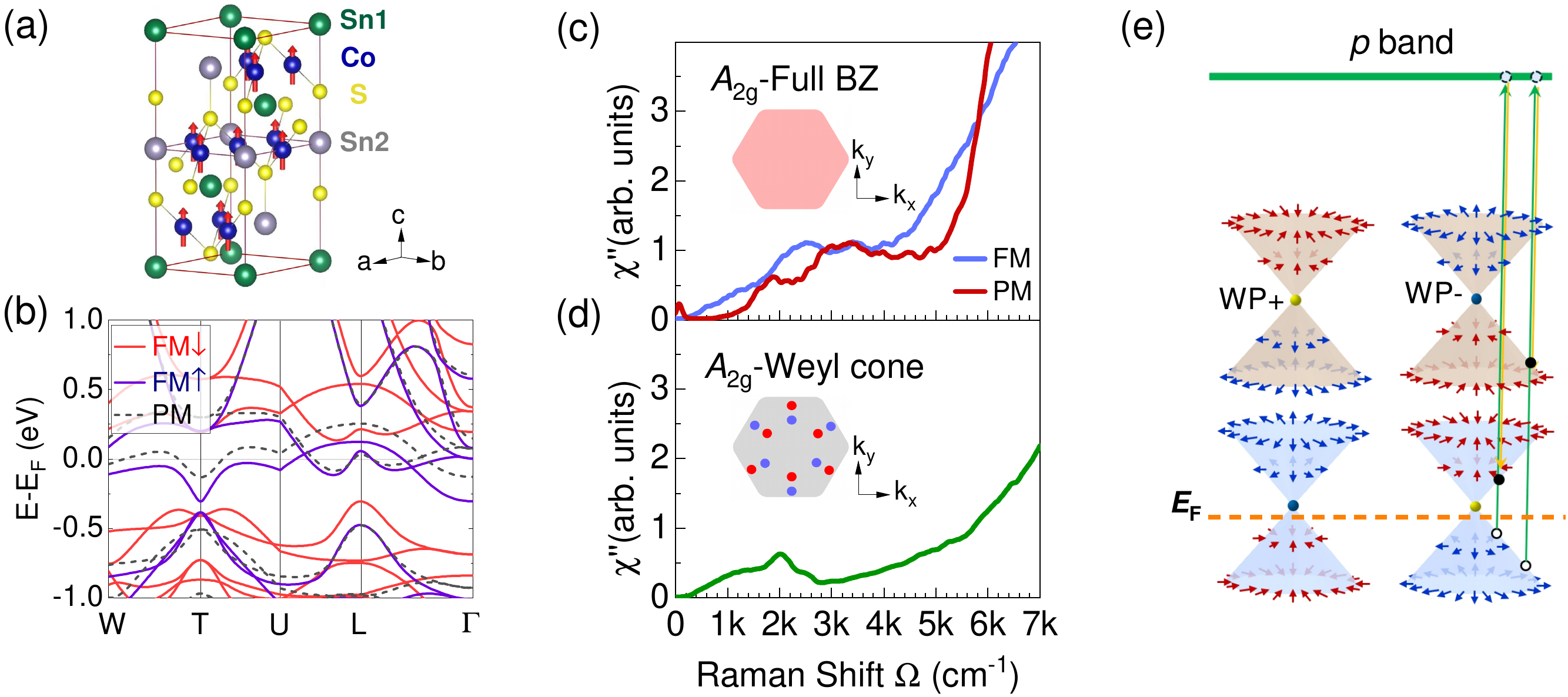}
  \caption{ \textbf{Simulated electronic Raman response of \CSS.} (a) Crystal structure of \CSS with the Kagome plane perpendicular to the $c$-axis. The ferromagnetically (FM) ordered spins are indicated. (b) Band structure for the paramagnetic (PM, dashes) and the FM (full lines) states. (c) and (d) Simulated \AZg Raman response using the Kramers-Heisenberg formula (see SI Section K). (c) Summation over all \textbf{k}-points in the Brillouin zone (BZ). (d) Summation over \textbf{k}-points near the Weyl cones covering only 0.8\% of the BZ per Weyl point. (e) Schematic of the resonant \AZg response in the FM state around the Weyl cones via Sb $p$-orbitals. The light blue and orange Weyl cones originate from the spin-up and spin-down dominated bands. The blue and red arrows on the cones indicate the spin-orbit locked spin textures. The schematic shows two pairs of Weyl points (WP). The Fermi level is set at the lower WPs. The \AZg response is attributed to excitations between bands of opposite chirality. }
	\label{fig:A2g}
\end{figure*}

\begin{figure*}[ht!]
  \centering
  \includegraphics[width=15cm]{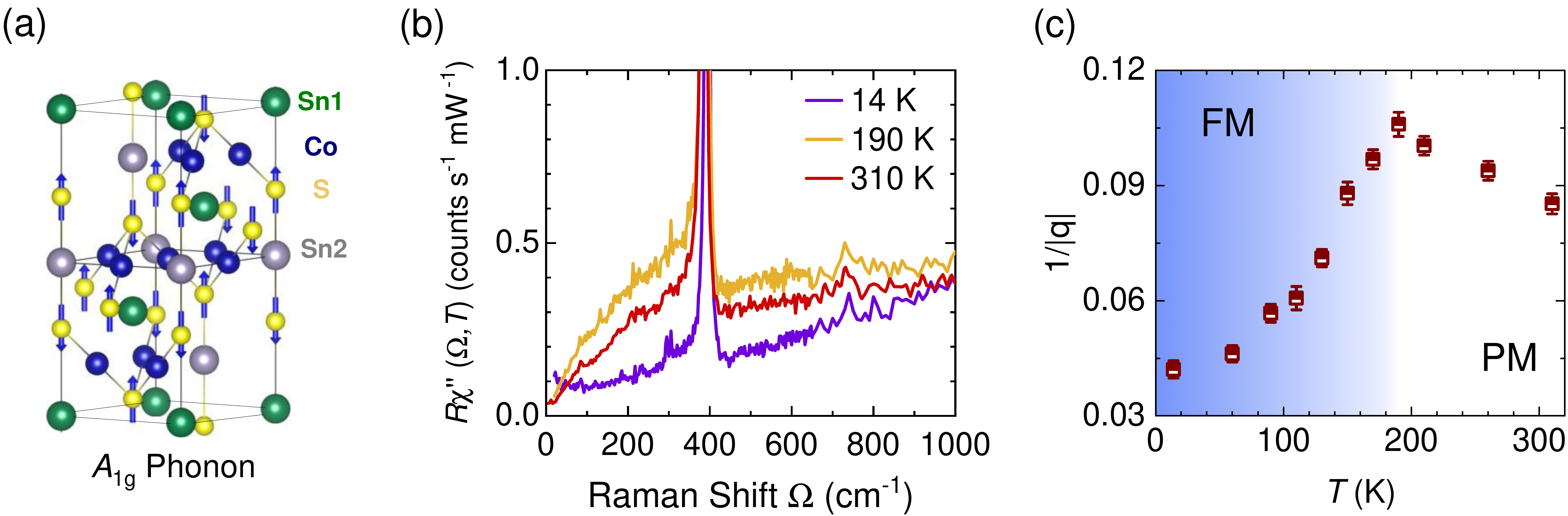}
   \caption{\textbf{\Alg phonon in \CSS.} (a) Eigenvector of the \Alg phonon. The sulfur atoms vibrate along the $c$-axis perpendicular to the Kagome plane. (b) \Alg spectra at temperatures as indicated. The phonon is superposed on a continuum that depends non-monotonically on temperature. Close to \Tc, the \Alg phonon line develops its strongest asymmetry. In the entire temperature range the phonon is best represented by a Fano function [Eq.~\ref{eq:fano}] as  shown explicitly in SI Section D.  (c)~Temperature dependence of the inverse asymmetry factor $1/|q|$ of the \Alg phonon extracted from the fits.
   }
	\label{fig:phonon_results}
\end{figure*}

\begin{figure}[ht!]
  \centering
  \includegraphics[width=8cm]{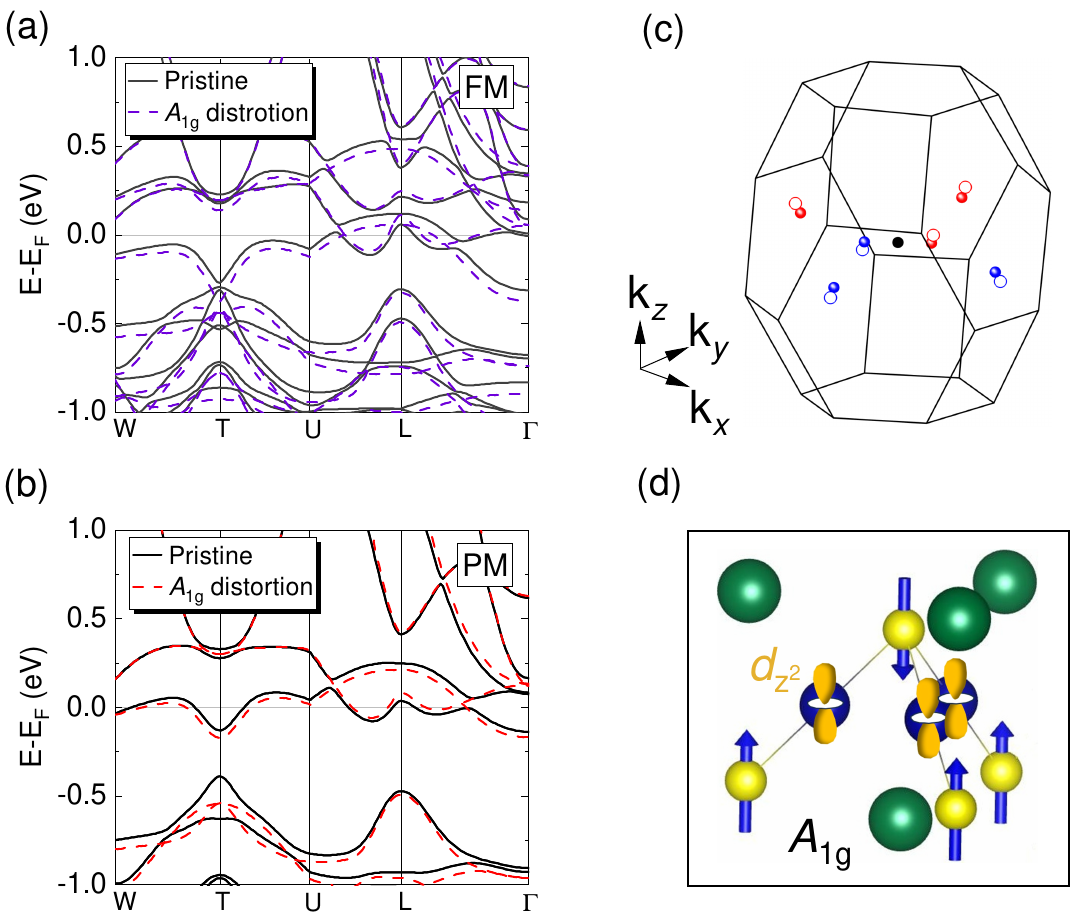}
  \caption{\textbf{Electronic structure of \CSS.} (a) Band structure with spin-orbit coupling for the FM state. The full and the dashed lines show the pristine bands and the effect of the \Alg distortion, respectively. (b) In the PM state the effect of the \Alg distortion is substantially weaker. (c) Effect of the \Alg distortion on the position of the Weyl points in the Brillouin zone. (d) Schematic of the electron-phonon coupling for the \Alg phonon. The $d_{z^2}$ orbitals are on top of the Co atoms (blue). Movement of the S atoms (yellow) toward or away from the Kagome plane, as is the case for the \Alg phonon, modulates the electric field across the $d_{z^2}$ orbitals. The two eigenvectors of the \Eg phonon are parallel to the Kagome plane, and thus, do not modulate the field across the plane, and the coupling is much weaker [see also Section \ref{sec:disc_phonon}].
  }
	\label{fig:Weyl}
\end{figure}

\subsection{Simulations for the continua}
\label{subsec:DFT}

The band structures of  the FM and PM states of \CSS can be found in Fig.~\ref{fig:A2g} (b). The PM band structure (dashed lines) has a spin-degenerate, nearly-flat band near the Fermi level, which splits into two distinct spin channels in the FM ground state (solid lines), separated by roughly 300~meV. Indeed, both Weyl bands shift relative to the Fermi level across the phase transition. This band shift may explain the spectral weight redistribution in the \Eg Raman spectra.

The suppression of intensity below 1200~\wn, most strongly developed in \Alg symmetry, can be reproduced only by including both intra- and interband transitions, an example of which is shown in SI Fig.~S17 (a) without considering selection rules. Since this paper's focus is on chiral excitations, we do not discuss this effect here, and instead refer to the SI.

To account for the selection rules, specifically the  \AZg channel and the projection of chiral excitations, we must calculate the resonant contributions to the Raman response using the Kramers-Heisenberg formula. The resulting theoretical spectra for the FM and PM states [see Fig.~S17 (b-e) in SI Section K] are roughly consistent with the experimental results: a peak at $\sim$ 2000~\wn followed by a steep increase at higher energies. This quantitatively reproduces the \Alg, \AZg, \Egl, and \Egz responses, and in particular, the large \AZg response, which is dominant at low energies for the FM state, having a peak at approximately 2400~\wn [see Fig.~\ref{fig:A2g} (c)].

We further investigate the resonant spectra using dense \textbf{k}-grids centered on the Weyl points, which are located near the Fermi energy, with their corresponding time-reversed partners sitting approximately 300~meV above the Fermi energy. Even though each of these \textbf{k}-grids projects only 0.8$\%$ of the total Brillouin zone volume, leading to appropriately weighted spectra, the low energy peak at approximately 2000~\wn still maintains about half the intensity of the peak that was obtained by integrating over the full Brillouin zone [see Fig.~\ref{fig:A2g} (c) and (d)]. Thus, we infer that excitations between the two Weyl bands close to the Weyl points are responsible for the anomalous \AZg electronic continuum. \\

\subsection{Phonons} \label{subsec:phonons}

Superimposed on the continua, there are two Raman-active phonons as shown in Fig.~\ref{fig:pure_symmetry} (a), (c), and (d). The degenerate \Eg phonon [(Fig.~\ref{fig:pure_symmetry}~(c) and (d)] (in-plane vibration of the S atoms)  exhibits an almost symmetric line shape, whereas the \Alg phonon [Fig.~\ref{fig:pure_symmetry}~(d)] (out-of-plane vibration of S atoms [see Fig.~\ref{fig:phonon_results}~(a)]) has an obviously asymmetric line shape above and around \Tc, consistent with previous Raman results \cite{Tanaka:2023}. The plots of the \Alg spectra, as displayed in Fig.~\ref{fig:phonon_results} (b), highlight the temperature dependence of the \Alg continuum, which was already visible in Fig.~\ref{fig:pure_symmetry}~(a), and the asymmetry of the phonon line shape, most prominently at 190~K close to the FM transition.
 In the limit $T\to0$ the line becomes nearly symmetric. Since the linewidth $\Gamma$ is much smaller than the resonance energy $\omega_{\rm ph}$ the usual (simplified) expression for the Fano resonance, valid close to $\omega_{\rm ph}$, 
\begin{equation}\label{eq:fano}
  A(\omega)=\frac{A_0}{|q^2+1|}\frac{(q+\varepsilon)^2}{1+\varepsilon^2}+c(\omega);
  ~~~\varepsilon=2\frac{\omega-\omega_{\rm ph}}{\Gamma},
\end{equation}
may be used to extract the asymmetry parameter $q$, which describes the coupling between the harmonic oscillator (phonon) and the continuous electronic excitations $c(\omega)$. Taking $|q|\to\infty$, or equivalently $1/|q|\to0$, recovers a Lorentzian phonon line shape. The electron-phonon coupling $\lambda_{\rm EPC}$ is only one factor contributing to $|q|$, whereas other factors are smooth functions of energy in the normal state of a metal that modify only the numerical value of $|q|$ \cite{Opel:1999}. For modeling purposes, $A_0$ is the oscillator strength of the phonon and a phenomenological expression is used for the continuum $c(\omega)$. The analysis of the phonon lines in the entire temperature range, as shown in detail in SI Section D, yields a strong variation of the asymmetry parameter $1/|q|$ for the \Alg phonon peaking at 190~K [Fig.~\ref{fig:phonon_results}~(c)] and a rapid reduction of the line width $\Gamma^{\rm A_{1g}}$ below \Tc as shown in Fig.~S3 of SI Section D. 

The energy and linewidth of the \Eg phonon do not exhibit unusual variations with temperature and may be described in terms of thermal expansion, according to the Gr\"uneisen theory, and anharmonic decay, respectively. Details in determining the related parameters are summarized in SI Section D. Energies, linewidths, phonon-phonon coupling constants, and the Gr{\"u}neisen constants for both phonons are listed in Table~II of the SI, together with the results of phonon calculations.  The calculations  indicate that the EPC strength is significantly higher for the \Alg than for the \Eg phonon, and that both figures are higher for the PM than the FM phase. The large EPC in the PM state explains why the Fano effect on the \Alg phonon survives above \Tc.

The band response to the phonon displacements is summarized in Fig.~\ref{fig:Weyl} and SI Section I. It is evident that the \Alg phonon elicits a much larger change in the band structure than the \Eg phonon in both the PM and FM states as shown in Fig.~S14.  The large band shift indicates that both the DOS  and the band response drive the increase in EPC strength.

The phonon-induced band response results in a shift of the Weyl points in reciprocal space as shown in Fig.~\ref{fig:Weyl} (c). A 0.02~{\AA} shift of the S atoms along the phonon eigenvectors moves the Weyl points by $8.0 \times 10^{-4}$ {\AA}$^{-1}$ and $6.8 \times 10^{-4}$ {\AA}$^{-1}$ for the two degenerate \Eg phonons, and by $5.0 \times 10^{-3}$ {\AA}$^{-1}$ for the \Alg phonon, nearly an order of magnitude more, as shown in detail in SI Section I, implying that the \Alg phonon couples more strongly to the Weyl fermions than the \Eg mode.

\section{Discussion}

\subsection{Electronic excitations}

Our Raman experiments revealed anti-symmetric \AZg excitations in the FM state, which are fundamentally associated with chiral excitations \cite{Ko:2010, Kung:2015, Riccardi:2016}. In our case, this is confirmed as an interband transition between the two pairs of spin-orbit-locked Weyl bands (see Fig. \ref{fig:A2g}(e)). Normally, spin-flip transitions are not allowed in non-resonant Raman processes. However, spin-orbit coupling breaks SU(2) symmetry near the Weyl points. In our Wannier orbital representation, the bands appear as superpositions of spin-up and spin-down states, allowing us to observe excitations between bands that are nominally of opposite spin when spin-orbit coupling is not considered. In other words, the spin-orbit coupling enables dipole transitions that would otherwise be forbidden. The dipole operator, as described in SI Section K, does not connect states of opposite spin. Instead, it connects the parallel-spin components of two states, provided the transition between the component orbitals is allowed for a given symmetry channel. The anti-symmetric excitation is ensured by the resonance process through the high-energy Sb $p$-band [see Fig.~\ref{fig:A2g}(e)]. Such an anti-symmetric continuum was previously reported from Dirac cone excitations in graphene \cite{Riccardi:2016}, and our experiments provide another typical example of chiral excitations from the Weyl cones.

The polarization dependence of the electronic spectra directly indicates the existence of selection rules. Normally, there are no coherence effects in a ferromagnet similar to those in a superconductor. Thus, the intensity redistribution is presumably a result of the reconstruction of the band structure alone. Indeed, spectral weight redistribution has been observed in various experiments, including optical conductivity \cite{Yang:2020, Xu:2020}, magneto-optic Kerr effect \cite{Okamura:2020}, and resonant inelastic X-ray scattering \cite{Abhishek:2022}. This redistribution is associated with the band shift induced by the ferromagnetic phase transition \cite{Yang:2020}, which has been directly observed in recent ARPES measurements \cite{Liu:2021}.

Our DFT simulations also demonstrate splitting between bands that host spin-up and spin-down electrons. If one assumes that interband transitions dominate the spectra, then, one may conclude that the energy differences between some bands around the Fermi surface increase below 1400~\wn, quenching the transitions available above the phase transition.

\subsection{Phonons}\label{sec:disc_phonon}

While we cannot fully clarify the origin the non-monotonic enhancement of EPC to the \Alg phonon from the Raman experiments and DFT calculations at this point, we do note that the DOS, which was obtained when accounting for the effective magnetic moment of the Co atom, does explain the increased EPC in the PM state. However, this dependence alone does not explain why the response is larger for the \Alg phonon than the \Eg phonon. To understand this dichotomy, we look at the eigenvectors.

Essentially, the EPC originates from the electrostatic potential across the Co plane which is modulated  by the displacement of the vibrating S atoms.  Specifically, the \Alg phonon in \CSS is  predominantly   a vibration of S atoms perpendicular to the Kagome plane [see Figs.~\ref{fig:phonon_results} (a) and \ref{fig:Weyl} (d)]. It induces density fluctuations of the Co $d$-electrons along the $c$-axis due to the asymmetric crystal field environment across the Kagome plane. This, in principle, would enhance EPC, in a similar fashion as the \Blg phonon in \YBCO (Ref.~\cite{Devereaux:1995}). Among the five Co~$d$ orbitals , the strictly in-plane $d_{xy}$ and $d_{x^2-y^2}$ orbitals couple more weakly to the out-of-plane S motion of the \Alg phonon than the $d_{xz}$, $d_{yz}$, and $d_{z^2}$ orbitals. Among them, the $d_{z^2}$ orbital dominates the EPC, since it contributes much more to the DOS near the Fermi level than the $d_{xz}$ and $d_{yz}$ orbitals (see Fig.~\ref{fig:Weyl}~(d) and SI Section~J for details). Furthermore, the DOS originating from the $d_{z^2}$ orbital in the PM state is much higher than that in the FM state thus explaining the reduced EPC in the FM state. In contrast, the motion of the S atoms is in-plane for the \Eg phonon, and the modulation of the electric field across the Kagome plane is negligible (SI Section~J). Consequently, the coupling to Co 3$d_{z^2}$ electrons is much weaker, substantially reducing the band shifts near the Fermi level. In a similar fashion, the motion of the Weyl points induced by the phonons is almost an order of magnitude larger for the \Alg than the \Eg phonon.

\section{Conclusions}

In summary, we studied the Raman spectra of single crystal samples of \CSS  as a function of polarization and temperature; and we provide a theoretical interpretation of the results. The symmetry-resolved electronic continua show a redistribution of spectral weight below \Tc, from below to above 1200\,cm$^{-1}$. The piled-up intensity peaks at approximately 2000\,cm$^{-1}$. Since a ferromagnet does not exhibit gap-like structures in a single band but rather a shift of the spin-up and spin-down bands with respect to the Fermi energy, we conclude that the redistribution of intensity below \Tc originates predominantly from band shifts. Further resolving the spectra in pure symmetries allows us to discern the \AZg chiral excitations originating from the Weyl bands.\\

 The two Raman-active phonon modes were identified to have \Alg and \Eg symmetry. The \Eg mode has a conventional Lorentzian shape. The \Alg  phonon exhibits an asymmetric Fano-type line shape, indicative of strong coupling to the electronic continuum. The asymmetry factor $1/|q|$ varies non-monotonically with temperature  and increases upon approaching \Tc from below and decreases again above.  Electrons in the $d_{z^2}$ orbital, among the Co 3$d$ electrons, are involved predominantly in EPC to the \Alg phonon through the crystal field gradient along the $c$-axis. These results indicate an evolution of  the electronic structure as a function of temperature along with the magnetic moment. The analysis of the EPC highlights the importance of the $d_{z^2}$ orbital for the low energy excitations in \CSS. The evolution of electronic continua, especially the chiral \AZg response, the EPC of \Alg phonon with temperature and magnetization, and the motion of the Weyl points in response to the \Alg displacement argue for an  entanglement of lattice structure, magnetic order, and topology. \\

\noindent\textbf{Acknowledgments}\\
We thank C.-J. Yi, R. Yang, D.-F. Liu, Y.-F. Xu, Y.-T. Sheng, Z.-D. Song for fruitful discussions. This work is supported by  the National Natural Science
Foundation of China (Grant No. 12474473), the National Key Basic Research Program of China
(Grants No. 2024YFB4007301 and No. 2024YFF0727100), the Deutsche Forschungsgemeinschaft (DFG) through the coordinated programme TRR80 (Projekt-ID 107745057) and projects HA2071/12-1 and -3. L.P. and R.H. were partially supported by the Bavaria-California Technology Center (BaCaTeC) under grant number A3 [2022-2]. G.H. would like to thank the Alexander von Humboldt Foundation for a research fellowship. D.J.
gratefully acknowledges funding of the Alexander von Humboldt foundation via a Feodor Lynen postdoctoral
fellowship. The work for sample synthesis and characterization was supported by grants from the Informatization Plan of the Chinese Academy of Sciences (CAS-WX2021SF-0102), and the Synergetic Extreme Condition User Facility (SECUF). Work at SLAC National Accelerator Laboratory and Stanford University (M.K., D.J., E.M.B, B.M., T.P.D.) was supported by the U.S. Department of Energy, Office of Basic Energy Sciences, Division of Materials Sciences and Engineering, under Contract No. DE-AC02-76SF00515. The computational results utilized the resources of the National Energy Research Scientific Computing Center (NERSC), a U.S. Department of Energy, Office of Science User Facility, using NERSC award BES-ERCAP0031424. \\

\noindent\textbf{Author contributions}\\
G.H. and R.H. conceived the project.
G.H., L. P., R.S., A.B. and D.J. performed the Raman measurements. G.H., M.K., L.P., J.S., T.P.D., and R.H. analysed the Raman data. Z.C.X. and Y.G.S. synthesised and characterised the samples. M.K., E.M.B., B.M. and T.P.D. performed
DFT calculations. G.H., M.K. and R.H. wrote the manuscript with comments
from all the authors.\\

\noindent\textbf{Competing interests}\\
The authors declare no competing interests.

\nocite{*}
\bibliography{refs} 

\end{document}